\documentclass[iop]{emulateapj}

\def\ltsima{$\; \buildrel < \over \sim \;$}
\def\simlt{\lower.5ex\hbox{\ltsima}}
\def\gtsima{$\; \buildrel > \over \sim \;$}
\def\simgt{\lower.5ex\hbox{\gtsima}}

\begin{document}

\title{Magnetically Controlled Accretion on the Classical T Tauri Stars GQ Lupi
       and TW Hydrae}

\author{Christopher M. Johns--Krull \& Wei Chen}
\affil{Department of Physics \& Astronomy, Rice University, 6100 Main St.
       MS-108, Houston, TX 77005, USA}
\email{cmj@rice.edu, wc2@rice.edu}

\author{Jeff A. Valenti}
\affil{Space Telescope Science Institute, 3700 San Martin Dr., Baltimore, MD
       21210, USA}
\email{valenti@stsci.edu}

\author{Sandra V. Jeffers}
\affil{Georg-August-Universit\"at, Institut f\"ur 
       Astrophysik, Friedrich-Hund-Platz 1, D-37077 G\"ottingen, Germany}
\email{jeffers@astro.physik.uni-goettingen.de}

\author{Nikolai E. Piskunov, Oleg Kochukhov, V. Makaganiuk, \& H.~C.~Stempels}
\affil{Department of Astronomy and Space Physics, Uppsala University, 751 20
       Uppsala, Sweden}
\email{piskunov@fysast.uu.se, Oleg.Kochukhov@fysast.uu.se, vitaly.makaganiuk@gmail.com, eric.stempels@fysast.uu.se}

\author{Frans Snik, Christoph Keller, \& M. Rodenhuis}
\affil{Sterrewacht Leiden, Leiden University, Niels Bohrweg 2, 2333 CA, 
       Leiden, The Netherlands}
\email{snik@strw.leidenuniv.nl, keller@strw.leidenuniv.nl, rodenhuis@strw.leidenuniv.nl}


\begin{abstract} 

We present high spectral resolution ($R\approx108,000$) Stokes $V$ polarimetry
of the Classical
T Tauri stars (CTTSs) GQ Lup and TW Hya obtained with the polarimetric upgrade
to the HARPS spectrometer on the ESO 3.6 m telescope.  We present data on
both photospheric lines and emission lines, concentrating our discussion on 
the polarization properties of the \ion{He}{1} emission lines at 5876 \AA\ and
6678 \AA.  The \ion{He}{1} lines in these CTTSs contain both narrow emission
cores, believed to come from near the accretion shock region on these stars,
and broad emission components which may come from either a wind or the
large scale magnetospheric accretion flow.  We detect strong polarization in
the narrow component of the two \ion{He}{1} emission lines in both stars.  We
observe a maximum implied field strength of $6.05 \pm 0.24$ kG in the 5876 \AA\
line of GQ Lup, making it the star with the highest field strength measured 
in this line for a CTTS.  We find field strengths in the two \ion{He}{1} lines
 that are consistent
with each other, in contrast to what has been reported in the literature on at
least one star.  We do not detect any polarization in the broad component
of the \ion{He}{1} lines on these stars, strengthening the conclusion that they
form over a substantially different volume relative the formation region
of the narrow component of the \ion{He}{1} lines.

\end{abstract}


\section{Introduction} 

T Tauri stars (TTSs) are young ($\simlt 10$ Myr), low mass 
($\simlt 2.5$ M$_\odot$) stars that have only recently emerged from their 
natal molecular cloud cores to become optically visible.  
These young, low mass stars 
are generally subdivided into categories such as classical and weak 
TTSs.  The designation of a classical TTS (CTTS) 
was originally based on a purely observational distinction: the equivalent
width of the H$\alpha$ emission line.  Classical TTSs are TTSs which
have an H$\alpha$ equivalent width $W_{eq}(H\alpha) > 10$ \AA\ as 
distinguished from the weak line TTSs (WTTSs) defined by
Herbig and Bell (1988); however, Bertout (1989) suggests
that a break point value of 5 \AA\ is more appropriate.  More recently,
investigators have tied the definition to the shape (width) of the H$\alpha$
line profile (e.g. White \& Basri 2003, Jayawardhana et al. 2003).  
Independent of the exact constraint imposed for defining a CTTS, this moniker
has become synonymous with a low mass pre-main star that is actively accreting
material from a circumstellar disk.  Indeed, the vast majority of stars
which fit the criteria for CTTSs show some kind of additional evidence
[e.g. inverse P-Cygni line profile shapes, optical veiling (see below),
infrared excess] indicative of disk accretion.  

It is now generally accepted that accretion of circumstellar disk material
onto the surface of a CTTS is controlled by a
strong stellar magnetic field (e.g. see review by Bouvier et al. 2007).
These magnetospheric accretion models assert that strong stellar magnetic 
fields truncate the inner disk, typically near the corotation radius, and 
channel the accreting disk material onto the stellar surface, most often 
at high stellar latitude  (Camenzind 1990;
K\"onigl 1991; Collier Cameron \& Campbell 1993; Shu et al. 1994; Paatz \& 
Camenzind 1996; Long, Romanova, \& Lovelace 2005).  More recent 
magnetohydrodynamic simulations find that outflows launched from
near the region in which the stellar field interacts with the surrounding 
accretion disk can also spin the star down to observed rotation
rates (e.g. Ferreira 2008; Romanova et al. 2009), though some recent work 
challenges the notion that these outflows can actually balance the spin-up accretion
torques in CTTS systems (e.g. Zanni \& Ferreira 2009).

Despite the successes of the magnetospheric accretion model, open issues remain.  Most current theoretical models
assume the stellar field is a magnetic dipole with the magnetic axis aligned
with the rotation axis.  However, recent spectropolarimetric measurements
show that the fields on TTSs are probably not dipolar
(Johns--Krull et al. 1999a; Valenti \& Johns--Krull 
2004; Daou, Johns--Krull, \& Valenti 2006; Yang, Johns--Krull, \& Valenti
2007; Donati et al. 2007, 2008, 2010a, Hussain et al. 2009).  Few studies of 
accretion onto CTTSs have taken into account non-dipole field 
geometries.  The earliest of these by Johns--Krull and Gafford (2002) found 
that abandoning the dipole assumption reconciled observed trends in the data
with model predictions; however, this study did not consider the torque 
balance on the star and whether an equilibrium rotation rate could actually be 
achieved.  Johns--Krull and 
Gafford (2002) argued that while the field on the stellar surface may be 
quite complex, the dipole component of the field should dominate at distance
from the star where the interaction with the disk is taking place.  This 
assumption appears to generally hold true in several recent studies (e.g. 
Johns--Krull \& Gafford 2002; Mohanty \& Shu 2008; Gregory et al. 2008; Long et al. 2008; Romanova et al.  2011; Cauley et al. 2012).  However, the complex nature of the field near the surface has
significant implications for the size of accretion hot spots, making them
smaller than would be predicted by pure dipole models (Mohanty \& Shu
2008; Gregory et al. 2008; Long et al. 2008); and also has important
consequences for disk truncation radii and the computation of the torque
balance on the star by the disk (Gregory et al. 2008; Long et al. 2008;
Romanova et al. 2011).  

Two approaches are generally used to measure magnetic fields on low mass
stars, both utilizing the Zeeman effect.  Magnetic fields can be measured
from the broadening of magnetically sensitive lines observed in intensity
spectra (e.g. Johns--Krull 2007; Yang et al. 2008).  This technique is
primarily sensitive to the magnetic field modulus, the unsigned value 
of the field weighted by the intensity distribution of the light emitted over
the visible surface of the star.  
While this method does not suffer
from flux cancellation due to regions of opposite polarity appearing on
the star, it does require that all non-magnetic broadening mechanisms 
be accurately accounted for in the observed spectra.  As a result, this
technique is primarily sensitive to relatively strong fields.  Observations
of circular polarization in Stokes $V$ spectra can be much more sensitive
to weak fields on the surface of stars; however, the Stokes $V$ signature is
sensitive only to the line of sight component of the magnetic field 
and the signal can be reduced significantly due to flux cancellation when 
opposite field polarities are observed simultaneously on the stellar surface.
Doppler shifts due to stellar rotation can reduce the degree of flux
cancellation that results, permitting Stokes $V$ signatures to be 
present even when the net flux weighted line of sight field integrated 
over the stellar surface (the net longitudinal magnetic field, $B_z$) is
zero.  Observations of time series of Stokes $V$ spectra can be used to 
track changes in the amount of net field visible on the star as it rotates,
ultimately allowing the large scale field of the star to be mapped using 
various tomographic imaging techniques (e.g. Donati et al. 2007 and references
therein; Kochukhov et al. 2004 and references therein).  

In addition to potentially mapping the surface field on accreting young
stars, information can be obtained on the large scale field controlling
the interaction of the star with its disk and the accretion flow by
measuring time series of Stokes $V$ profiles in emission lines formed in
the accretion flow and shock.  
The first accretion line for which circular polarization
was detected is the \ion{He}{1} line at 5876 \AA\ (Johns--Krull et al.
1999a), and time series of the polarization variations in this line
have been used to estimate the latitude of accretion spots on several CTTSs
(e.g. Valenti \& Johns--Krull 2004; Yang et al. 2007; 
Donati et al. 2008, 2010b, 2011a, 2011b).  This 
line is observed in most
CTTSs and is often found to be composed of two components: a narrow
core component and a broad component extending out to several hundred
km s$^{-1}$ (e.g. Edwards et al. 1994; Batalha et al. 1996; Alencar \& 
Basri 2000).  Based on the similarity in shape between the observed line
profiles of some CTTSs and model profiles calculated in the context of
magnetospheric accretion, Hartmann et al. (1994) suggested 
that the \ion{He}{1} 5876 \AA\ line (broad and narrow components) might 
form throughout the accretion flow, with the narrow component primarily
coming from the lower velocity regions near the disk truncation point.
Beristain et al. (2001) instead argue that the narrow core of the 
\ion{He}{1} line arises in decelerating post-shock gas on the stellar
surface at the base of the accretion footpoints.  
Beristain et al. (2001) argue that the broad component
observed in many CTTSs has a dual origin in the magnetospheric flow and
in a high velocity wind in the most strongly accreting stars.  

The strong, ordered fields observed in the narrow component of
this line component ($\sim 2.5$ kG, e.g. Johns--Krull 
et al. 1999a) argue for a formation region close to the stellar surface 
instead of several stellar radii above the star where the field interacts
with the disk.  The \ion{He}{1} 5876 \AA\ arises from a triplet state
and is composed of several closely spaced lines.  The \ion{He}{1} 6678 \AA\
line arises from the analogous singlet state, and is observed in many
CTTSs as well where it displays both broad and narrow components (see 
Beristain et al. 2001).  Based on the strong similarity in their kinematic
properties and the measured triplet-singlet flux ratio, Beristain et al.
(2001) conclude that the narrow component of both \ion{He}{1} lines forms
in the post-shock gas.  On the other hand, this picture is complicated
by the observation of Donati et al. (2008) that the 6678 \AA\ line 
consistently shows a longitudinal magnetic field strength approximately twice
that of the 5876 \AA\ line in the CTTS BP Tau whose \ion{He}{1} lines
are dominated by a narrow component (Edwards et al. 1994, Batalha et al. 
1996, Beristain et al. 2001).  This is a surprising observation since
models of accretion shocks on CTTSs find that the thickness of the post-shock
region is typically $10^5 - 10^6$ cm (Calvet \& Gullbring 1998; Lamzin
1998) which is a small fraction ($\simlt 10^{-5}$) of a stellar
radius.  It would be surprising if the stellar magnetic field strength
varied so strongly with depth, suggesting then that perhaps the two
\ion{He}{1} lines do not trace the same regions on the stellar surface.

To better clarify the magnetic field properties of accretion related lines,
more spectropolarimetric observations of CTTSs, including those with
substantial broad components to their \ion{He}{1} lines, are needed.
Here, we report new observations of two CTTSs (GQ Lup and TW Hya) using
the newly commissioned polarimeter operating with the HARPS spectrograph
on the ESO 3.6 m telescope at La Silla.  TW Hya is a K7 CTTS and a member
of the loose TW Hydrae association (Kastner et al. 1997).  The {\it Hipparcos}
parallax for TW Hya implies a distance of $56 \pm 7$ pc (Wichmann et al.
1998), making it the closest CTTS to the Earth.  Based on its placement in 
the HR diagram, the age of TW Hya is estimated to be 10 Myr (Webb et al. 1999,
Donati et al. 2011b).  Setiawan et al. (2008) claimed the detection of a
$\sim 10$ M$_{Jup}$ planet in a very close orbit around this CTTS, making
TW Hya an important benchmark constraining the timescale of planet 
formation.  Hu\'elamo et al. (2008) instead suggest that the observed
radial velocity variations which signal the presence of the planet are
in fact caused by large starspots on the surface of TW Hya.  As a result,
there is great interest in knowing as much about this star as possible.
In addition, TW Hya is still accreting material from its circumstellar disk
and is observed at a low inclination ($i \sim 18^\circ$, Alencar \&
Batalha 2002), making it an excellent object for studying magnetically
controlled accretion onto young stars.  The magnetic properties of TW Hya
have been investigated a number of times previously (Yang et al. 2005, 2007;
Donati et al. 2011b).  GQ Lup is also a K7 CTTS, and has also recently
come under a great deal of scrutiny as the result of a claimed planetary
mass companion.  Neuh\"auser et al. (2005) discovered an infrared 
companion at a separation of $\sim 0.^{\prime\prime}7$ (corresponding to
$\sim 100$ AU at a distance of 150 pc).  Based on their infrared photometry
and K band spectra, Neuh\"auser et al. (2005) constrained the mass of
GQ Lup B to be between $1-42$ M$_{\rm Jup}$, placing it possibly in the
planet regime.  More recent spectroscopic studies have favored the upper
end of this range, suggesting the companion is more likely a
brown dwarf (Mugrauer \& Neuh\"auser 2005; Guenther et al. 2005; 
McElwain et al. 2007;
Seifahrt et al. 2007; Marois et al. 2007; Neuh\"auser et al. 2008).  The
formation of such an object presents challenges to theories of companion
formation in a disk, and has sparked continued study of this system to
better pin down the properties of both of its members.  GQ Lup is known to
show clear signs of variable accretion (Batalha et al. 2001), making it a
good target to study the role of magnetic fields in the accretion process.
To our knowledge, no studies of the magnetic properties of GQ Lup exist
to date.
In \S 2 we describe our observations and data reduction.  The magnetic field 
analysis and results are described in \S3, and in \S 4 we discuss the 
implications of our findings.

\section{Observations and Data Reduction}

All spectra reported here were obtained at the ESO 3.6 m telescope on La 
Silla using the newly comissioned polarimeter, HARPSpol (Snik et al. 2008,
2011; Piskunov et al. 2011), mounted in front of the fibers feeding the HARPS 
spectrometer (Mayor et al. 2003).  While HARPSpol can also record Stokes 
$Q$ and $U$ spectra, for the observations reported here, only Stokes $V$ 
spectra were obtained.  As mentioned above, linear polarization in both the
lines and the continuum can result from scattering off a circumstellar disk
(e.g. Vink et al. 2005); however, the action of a disk does not typically
produce circular polarization in either the lines or the continuum.
Here, we will focus only on Stokes $V$ in the lines measured 
relative the continuum which is assumed to not be circularly polarized.  With
this instrumental setup, each exposure simultaneously records the right and 
left circularly polarized components of the $R = 108,000$ spectrum.  These two
components of the echelle spectrum are interleaved, such that two copies
of each echelle order are present on the two $2148 \times 4096$ CCD arrays
(one for the blue portion of the spectrum and one for the red).
The two polarized components of each order are separated by $\sim 16$
pixels in the cross dispersion direction on the array, while each spectral
trace is $\sim 3.5$ pixels wide (FWHM) in the cross dispersion direction.
Each observation of a star reported here actually consists of 4 separate
observations of the star, with the angle of the quarter waveplate in the
polarimeter advanced by 90$^\circ$ between the exposures.  The result of 
this is to interchange the sense of circular polarization in the two beams.
This gives substantial redundancy in the analysis which allows us to remove 
most potential sources of spurious polarization due to uncalibrated 
transmission and gain differences in the two beams.  As described below,
we use the ``ratio" method to combine the spectra from these interchanged
beams in order to form Stokes $I$ and $V$ spectra that are largely free
of these potential spurious signals (e.g. Donati et al. 1997; Bagnulo et
al. 2009).
All spectra were obtained on the nights 29 April 2010 through 2 May 2010,
with one night (1 May) lost due to weather.
Table \ref{obslog} gives a complete table of the stellar observations 
reported here.  Included in the Table are continuum signal-to-noise estimates
near the two \ion{He}{1} emission lines studied here as well as the emission
equivalent widths of these two lines.  Also reported is the veiling found
near the \ion{He}{1} 6678 \AA\ line as discussed below.
Along with spectra of GQ Lup and TW Hya, a spectrum of 
the weakly accreting TTS V2129 Oph was
also obtained and is used in the analysis of the \ion{He}{1} lines on the
other stars.  In addition to stellar spectra, standard calibration
observations were obtained including bias frames, spectra of a Thorium
Argon lamp for wavelength calibration, and spectra of an incandescent 
lamp for the purpose of flat fielding.  The calibration spectra were obtained
with the polarimeter in front of the fibers.

All spectra were reduced with the REDUCE package of IDL echelle reduction
routines (Piskunov \& Valenti 2002) which builds on the data reduction 
procedures described by Valenti (1994) and Hinkle et al. (2000).  The 
reduction procedure is quite standard and includes bias subtraction, flat 
fielding by a normalized flat spectrum, scattered light subtraction, and 
optimal extraction of the spectrum.  The blaze function of the echelle
spectrometer is removed to first order by dividing the observed stellar
spectra by an extracted spectrum of the flat lamp.  Final continuum 
normalization was accomplished by fitting a 2nd order polynomial to the
blaze corrected spectra in the regions around the lines of interest for
this study.  Special care was taken to apply a consistent continuum 
normalization procedure to the spectra extracted from all four sub-exposures. 
Occasional small difference in normalization of the two orthogonal spectra 
are compensated by using the ``ratio" method (e.g. Bagnulo et al. 2009, and
below) to combine the right and left 
circularly polarized components.  The wavelength solution
for each polarization component was determined by fitting a 
two-dimensional polynomial to $n\lambda$ as function of pixel and order 
number, $n$, for approximately 1000 extracted thorium lines observed from the 
internal lamp assembly.  The resolution as determined by the median FWHM
of these thorium lines was $R = 107,660$.

As mentioned above, each subexposure obtained of a given star contains both 
the right and left circularly polarized component of the spectrum.  In order
to get a final measurement of the mean longitudinal magnetic field, $B_z$,
these individual measurements of the two circular polarization components
must be combined in some way.  We used the ``ratio" method (e.g. Bagnulo et 
al. 2009, Donati et al. 1997) to combined the right and left circularly 
polarized components of the spectra form the Stokes $V$ spectrum as well as 
a null spectrum, with each being renormalized to the continuum intensity.  We 
also added all the components together to form the Stokes $I$ spectrum.  With
Stokes $V$ and $I$ determined, the continuum normalized right-hand circularly
polarized (RCP) component of the spectrum is then $R = I + V$ and the 
continuum normalized left-hand circularly polarized (LCP) component of the
spectrum is $L = I - V$.  Computing these from $I$ and $V$ in this way 
ensures both circular polarization states have been normalized to the same 
continuum.

\section{Analysis}

\subsection{\ion{He}{1} Line Equivalent Widths}

Table \ref{obslog} gives the equivalent width of the two \ion{He}{1} lines
studied here for all our target stars.  As mentioned before, previous
investigators have noted that these lines often appear to have two
distinct components (e.g. Batalha et al. 1996; Beristain et al. 2001):
a narrow component (NC) and broad component (BC).  It is thought that the
two components may form in different physical regions of the accretion
flow onto CTTSs (Beristain et al. 2001) and their polarization properties 
also appear to be different with the narrow compnent showing significantly
stronger polarization (Daou et al. 2006; Donati et al. 2011b).  We
therefore report the equivalent width of the NC and the BC separately for
the two \ion{He}{1} lines, the sum giving the total line equivalent width.  
Decomposing the lines in this manner requires some assumptions to be made 
about how to separate the two components.  Since the NC often 
appears asymmetric (e.g. Figure 1) with a very steep blue edge and shallower 
red edge, Gaussian fitting to the lines requires particular choices to be
made on just how to do the analysis.  For example, Batalha et al. (1996)
define (by eye) a region outside the NC and fit a single Gaussian to the
resulting BC and subtract it off in order to measure the NC equivalent width.
Another procedure is to fit the entire line with multiple Gaussians and use
the resulting fit parameters to estimate component properties (e.g. Alencar
\& Basri 2000).  The resulting equivalent with of the various features then
depedns at some level on how one chooses to do the analysis.  This is 
illustrated in Figure 1.  The top panel shows the \ion{He}{1} 5876 \AA\ line
of TW Hya from the first night.  The smooth solid curve shows a line profile
fit employing 3 Gaussian components.  The dash-dot line shows a fit using
only two Gaussian components.  There is a clear difference in the two fits 
(the 3 Gaussian fit uses two Gaussians to fit the NC which is not really
Gaussian as mentioned above).

The bottom panel of Figure 1 zooms in on the line to show the recovered BC
profiles.  The BC from the 3 Gaussian fit is shown in the smooth solid line
and that from the two Gaussian fit is shown in the dash-dot line.  Also shown
is a BC fit (dash-triple dot line) following Batalha et al. (1996) where a 
single Gaussian is used to fit the region outside the NC.  Finally, the solid
straight line connecting the two large squares shows a by eye estimate of the 
point on both the blue and red side of the line (as seen in Stokes $I$) where
the NC and BC join with a linear interpolation between these points to 
define the separation of the NC and the BC which can be used to separately 
determine their equivalent widths.  This then gives 4 different ways to
estimate the equivalent width of the BC (and also the NC).  The two extremes
for the BC equivalent width are the single Gaussian fit (1.193 \AA) 
following Batalha et al. and that (1.089 \AA) from the two Gaussian fit,
corresponding to a difference of 9\%.  Clearly, none of the Gaussian fits
exactly follow the red side of the BC, so it is impossible to predict 
just what this component does under the NC.  Given this uncertainty and
the fact that using the linear interpolation between the blue and red sides
of the apparent boundary between the NC and BC gives equivalent width values
between the two extremes, we choose to use this method to separate both
components and measuure the equivalent widths reported in Table \ref{obslog}.
We note that the BC equivalent width for the profile shown in Figure 1
computed this way differs from that resulting from the 3 Gaussian fit by
only 4.9\%.  We therefore estimate that the systematic uncetainty resulting
from the choice of just how to separate the two components likely leads to
a 5\% uncertainty in the reported equivalent widths which is not included
in the Table.  In most cases, the boundary between the BC and the NC is clear
and repeated measurements with slightly different choices yield results with 
a difference less than $1 \sigma$ for the quoted uncertainties.  There are 
a few cases where the boundary between the NC and the BC, or the BC and the 
continuum, are less clear and we repeated the measurements with a larger 
distinction in our choices of these points.  These are noted
Table \ref{obslog} and we use our different measurement trials to
estimate the the equivalent width uncertainty for these profiles.
For the other measurements, the uncertainties are computed by propagating
the uncertainties in the observed spectra.

\subsection{The Photospheric Mean Longitudinal Field}

For each of the T Tauri stars, we measured the photospheric $B_z$ using 
approximately 40 magnetically sensitive absorption lines (Table 2), which 
form primarily over the portions of the stellar surface that are at 
photospheric temperatures.  These lines may have relatively little 
contribution from the cool spots that are likely present on these stars.
Due to the wavelength
dependence of the Zeeman effect and the fact that the signal-to-noise
ratio achieved in the observations of these late-type stars is considerably
higher in the red regions of spectrum, we focus the analysis here only on 
the spectra from the red CCD of HARPSpol.  Lines for the analysis were selected 
by visual inspection of all the orders on the red CCD recorded with HARPS.
Lines were deemed good for the analysis if they appeared relatively strong
(central depth $\simgt 0.15$) in the observed spectrum (though most
were considerably stronger), appeared free of blending by other
photospheric lines, and were not contaminated by telluric absorption.
Lines passing these criteria were then checked in the Vienna
Atomic Line Database (VALD, Kupka et al. 1999, 2000) and if they are present
and have a value for the effective Land\'e $g$-factor, the line was used
in the analysis.  In a few cases, the VALD data indicated that an
apparently good line is actually a very close blend of two lines.  In this
case, we used the line but estimated a new effective Land\'e $g$-factor 
by calculating the weighted mean of the effective Land\'e $g$-factors
of the lines in the blend.  The weights used are the central depth of each 
component line as predicted by VALD for the atmospheric parameters typical
of K7 TTS ($T_{eff} = 4000$ K, log$g = 3.5$).  The initial line list was
constructed using a visual examination of the spectrum of GQ Lup obtained
on 29 April 2010.  For the other TTSs some lines were affected by
blending with telluric absorption or by strong cosmic ray hits (as is
also the case for later observations of GQ Lup).  In these cases, the
lines were not included in the determination of the photospheric $B_z$
values.  Lines so affected are noted in Table 2.

Once the line list was determined, the mean longitudinal magnetic
field, $B_z$, can be estimated by measuring the wavelength shift 
of each line, $\Delta\lambda = \lambda_R - \lambda_L$, where $\lambda_R$ is 
the wavelength of the line observed in the RCP component of the spectrum 
and $\lambda_L$ is the wavelength measured in the LCP component of the 
spectrum (Babcock 1962).  The shift of the line observed in the two 
polarization states is related to $B_z$ by
$$\Delta\lambda = 2{e \over 4\pi m_ec^2} \lambda^2 g_{\rm{eff}} B_z
                = 9.34 \times 10^{-7} \lambda^2 g_{\rm{eff}} B_z \,\,\,\,\,\,
                  \rm m\mbox{\AA}\,\eqno(1)$$
where $g_{\rm{eff}}$ is the effective Land\'e $g$-factor of the
transition, $B_z$ is the strength of the mean longitudinal
magnetic field in kilogauss, and $\lambda$ is the wavelength of the
transition in Angstroms (Babcock 1962; see also Mathys 1989, 1991).
In order to measure the wavelength shift, 
$\Delta\lambda$, we measured the wavelength of each line of interest
in the two circular polarization components using the so-called center of 
gravity method (e.g. Mathys 1989, 1991; Plachinda \& Tarasova 1999).  
This method for estimating $B_z$ is mathematically equivalent to estimating 
$B_z$ from the first order moment of the Stokes $V$ profile, assuming the
same integration limits are used in the two methods (e.g. Borra \&
Vaughn 1977; Mathys 1989; Landi Degl'Innocenti 2004).
Determining $B_z$ requires knowledge of the effective
Land\'e $g$-factor, $g_{\rm eff}$, for the transition.  The weights for
individual $\pi$ and $\sigma$ components of a given spectral line which
go into the definition of $g_{\rm eff}$ assume an optically thin line, 
so equation (1) is only approximately true in the case of moderately strong 
(saturated) photospheric lines.  As a result, using equation (1) is strictly 
valid only for weak lines, but give good results for real spectral lines 
(Mathys 1991).  For each measurement of the center of gravity of a spectral
line, we used the locally measured signal-to-noise in the spectrum to 
estimate the uncertainty in the spectrum and then used standard error
propagation to find the uncertainty in the line shift bewteen the two 
polarization states and the implied $B_z$.  Our final estimate of $B_z$ is 
a weighted mean of the individual line estimates, and these means are 
reported in Table 3.  

We repeated the measurements of the line wavelengths in the two polarization
components and the resulting final value of $B_z$ a number of times, making
slightly different choices on integration limits for the center of gravity 
estimate of the wavelength of each line.  However, for each trial we always
used the same integration limits for a given line when analyzing the RCP and 
LCP components of the spectra.  In each case, we achieved consistent results
within our quoted uncertainties.  Generally, we divided our trials in
two groups.  In the first case, we choose integration limits very close
to where the lines appear to reach the local continuum.  This was
primarily done as an effort to exclude any potential weak line blends 
that might appear as a small distortion in the line wings.  In the second 
group, we choose integration limits clearly out in the local continuum, but 
which in some cases likely included some weak line blends.  In many cases,
choosing the wider limits produced higher values of $B_z$, though in some
cases the measured field went down.  On average, the wider bins resulted
in fields stronger by $\sim 35$\%.  Choosing the wider limits does generally
result in somewhat larger uncertainty estimates, typically
by a factor of 1.8.  Again, the two groups of results are consistent within
these uncertainties (the difference typically being $\sim 1.5\sigma$).  In 
Table 3 we quote the values for the wider integration limits with their 
correspondingly greater uncertainties.  

Examining the photospheric fields and their uncertainties as reported in 
Table 3, apparently significant fields are found on both TW Hya and GQ Lup
each night they were observed. However, the value of $B_z$ found on V2129 
Oph is less than $1\sigma$ and does not represent a real detection.  As a 
test of our measurement techniques and in order to gain confidence in our
uncertainty estimates, two different null tests were performed on the
observations of each target.  Each test should return a measured value of
$B_z = 0$, thus serving to test how accurately we can recover this value
and whether the uncertainties are realistically estimated.
As described above, the spectra from the 4 subexposures were also combined
in such a way as to produce a null Stokes $V$ profile (e.g. Donati et al.
1997; Bagnulo et al. 2009), which can be used to calculate null RCP and LCP
spectra.  We analyzed these null spectra in
the same as described above.  This has the advantage of using exactly
the same lines as used to measure $B_z$, with exactly the same wavelength 
limits for computing the shift of each line, but with data combined in a 
different way than in the real measurement.  The second null test we
employed used 15 strong telluric lines from the atmospheric B band of
oxygen between 6883 - 6910 \AA.  Since these lines should show no
significant polarization of their own, we can analyze them in exactly the 
same fashion as we do the stellar lines from which we derive $B_z$ (that is 
we take $\Delta\lambda = \lambda_R - \lambda_L$ as done for the stellar 
measurements) combining the data from the sub-exposures in exactly the 
same way as done for the stellar measurements.  
In order to translate the measured shifts to a value of
$B_z$ we assign a Land\'e $g$-factor to each telluric line equal to the
mean value of the photospheric lines used for the given observation.  The
weights are the uncertainty on the value of $B_z$ derived from each of
the photospheric lines.  These null test field values are also reported in 
Table 3.

\subsection{$B_z$ in the Accretion Shock Emission}

As described above, Johns--Krull et al. (1999a) discovered that the 
\ion{He}{1} 5876 \AA\ emission line can be circularly polarized in spectra 
of CTTS, implying coherent magnetic fields at the footpoints of
accretion columns.  They measured $B_z = 2.46 \pm 0.12$ kG for BP Tau.
Since this original discovery, polarization in the \ion{He}{1} 5876 \AA\
emission line has now been reported for a number of CTTSs by a number
of investigators (Valenti \& Johns--Krull 2004; Symington et al.
2005; Smirnov et al. 2004; Yang et al. 2007; Donati et al. 2007, 2008,
2010b; Chen \& Johns--Krull 2011).  Polarization has since been 
detected in other emission lines (notably \ion{He}{1} 6678 \AA\ and the
\ion{Ca}{2} IRT lines) thought to be associated with accretion shock
emission as well (Yang et al. 2007; Donati et al. 2007, 2008, 2010b; Chen \&
Johns--Krull 2011).  The spectra obtained here contain both the 
\ion{He}{1} 5876 \AA\ and 6678 \AA\ lines, so we analyze them with a
focus on trying to see how well the fields derived from each line
agree with one another as this could provide clues to the location and
geometry of the accretion shocks on the star.  As mentioned earlier,
the 5876 \AA\ line of \ion{He}{1} is composed of several components
(6) which are closely spaced in wavelength, and as a result is subject
to the Paschen-Back effect (e.g. Yang et al. 2007; Asensio Ramos et al. 2008).
Therefore, the exact splitting pattern of the lines can vary 
considerably, depending on the strength of the magnetic field.  However,
since most of the level crossing and merging has occurred by the time
the local field strength reaches 2 kG (e.g. Asensio Ramos et al. 2008),
the \ion{He}{1} 5876 \AA\ line should be in or very close to the complete
Paschen-Back limit given the field strengths we recover below.  As a result,
we set $g_{eff} = 1.0$ for this line as done in Yang et al. (2007).  We can
then test the validity of this assumption once we have our field measurements.

Figure 2 shows the right- and left-circularly polarized components of
the \ion{He}{1} 5876 \AA\ emission lines of GQ Lup and TW Hya as observed 
on 29 April 2010.  Also shown in the figure are the Stokes $V$ profiles of
the lines.  Nearby photospheric absorption lines are also seen in each
star.  Since the polarization signal in the photosphere is quite weak due 
to the low value of $B_z$ present there (Table 3), these individual 
photospheric absorption lines do not show obvious polarization.  They do 
serve to show that the two polarization components are well aligned in 
wavelength, so that the obvious shift of the \ion{He}{1} emission line 
between the RCP and LCP components indicates a very strong field in the 
line formation region.  

In order to measure the value of $B_z$ in the line formation region of
the \ion{He}{1} line we again use the center of gravity technique to
measure the wavelength shift of the line as observed in the two polarization
components (RCP and LCP).  Our measurements of $B_z$ and 
its uncertainty for both \ion{He}{1} emission lines are reported in Table 3 
for each star on each night.  When measuring the field in the \ion{He}{1} 
formation region, care must be given when selecting the wavelength limits for
determing the center of gravity wavelength of the line and also in deciding
how to separate the narrow (NC) and broad (BC) components of the line 
(\S 3.1).  Looking at the \ion{He}{1} 5876 \AA\
line of TW Hya in Figures 1 and 2, polarization is clearly 
seen in the NC of the line, but is not obviously apparent
in the BC extending off to the red side of the line.  As
a result, we focussed in on the NC of the line when 
measuring the value of $B_z$ in the 5876 \AA\ line of \ion{He}{1}.

In addition to the specific wavelength region chosen, care must also be
taken when defining the local continuum to be used when measuring the 
center of gravity wavelength for the line.  The reason for this is that
the center of gravity technique (as well as the first moment of Stokes $V$)
is an intensity weighted mean wavelength,
where the intensity used is that above the continuum in the case of an
emission line.  In the case of the \ion{He}{1} 5876 \AA\ emission line, 
the NC of the line sits on top of a BC in
many cases as discussed above.  In this case, significantly different results
are obtained if the stellar continuum is used compared to what is
obtained if a somewhat higher continuum defined by the BC 
is used.  We proceed in an effort to isolate the emission from the NC 
and measure $B_z$ in this component.
Interpreting the NC of the \ion{He}{1} emission
as an excess line emitted from a distinct region that adds its emission to
that from both the stellar continuum and the BC of the
\ion{He}{1} emission, these additional sources should be subtracted off
when measuring the center of gravity wavelength of the NC 
of the emission line.  In \S 3.1 we described several methods of separating
the NC and BC when measuring their equivalent width, showing that each
method is subject to certain biases, but that the resulting systematic 
differences were small ($\sim 5$\% for the method we settle on).  We
used the same methods to remove the BC from the line (each BC is shown
in the bottom of Figure 1) and measured the resulting $B_z$.  For the
profile shown in Figure 1, using the single Gaussian fitted to the
region outside the NC gave the largest magnitude field ($-2.57 \pm 0.11$ kG),
while the two Gaussian fit and the linear interpolation both gave
$-2.38 \pm 0.10$ kG, and the 3 Gaussian fit gave $-2.40 \pm 0.11$ kG.
All 4 methods give results consistent to within 2$\sigma$ and three of
the 4 differ by 0.02 kG or less.  As in the case for the equivalent widths
above, we again adopt the linear interpolation under the NC as the way
of removing the BC and use the resulting NC to determine the $B_z$ values
given in Table 3.

As mentioned earlier, there have been reports in the literature that
the \ion{He}{1} 6678 \AA\ line shows substantially stronger polarization
than the 5876 \AA\ indicative of a stronger local field in the line 
formation region of this more optically thin line.  We looked for evidence 
of this effect in our stars by analyzing the the 6678 \AA\ emission line in
all 3 of the T Tauri stars oberved here in a way similar to how the
the 5876 \AA\ line is analyzed.  However, it became immediately apparent
that additional care needs to be taken when analyzing the 6678 \AA\ line.
Figure 3 illustrates the situation for GQ Lup with the spectrum obtained
on the first night of our observing run.  Shown in the figure is the
right and left circularly polarized components of the spectrum, along
with the Stokes $V$ profile.  Unlike the 5876 \AA\ line where the right and
left circularly polarized components of the emission line have essentially
the same shape, only shifted, the 6678 \AA\ line of GQ Lup has a 
different shape in the two circular polarization components.  On the other
hand, the Stokes $V$ profile for the two \ion{He}{1} lines looks quite similar.
The reason this is the case is the presence of an \ion{Fe}{1} photospheric
absorption line at the same wavelength of the 6678 \AA\ line which can be
seen in the spectrum of V2129 Oph (taken on night 2) which is also shown in
Figure 4.  In V2129 Oph, the \ion{He}{1} 6678 \AA\ line shows no obvious 
emission, but it turns out there is some weak emission in this line in V2129
Oph that partially fills in the photospheric absorption line and produces a 
clear Stokes $V$ signature which is illustrated in Figure 4 and discussed
below.

As described above, the formation region for the narrow components of
these \ion{He}{1} lines is thought to be at the base of the accretion
columns where material from the disk is raining down onto the star.  In
addition to producing some emission lines, these accretion footpoints 
are believed to be the source of the optical continuum veiling seen in
most CTTSs (e.g. Basri \& Batalha 1990; Hartigan et al. 1991; Valenti et al.
1993; Gullbring et al. 1998).  This extra line and continuum emission region
effectively blocks some small portion of the stellar surface and the light
it emits, adding its own emission on top of the stellar spectrum coming 
from the non-accreting regions of the star (e.g. Cavet \& Gullbring 1998).
The spectrum of the excess can then be studied by subtracting an 
appropritely scaled (veiled) stellar template spectrum from the observed
CTTS spectrum (e.g. Hartigan et al. 1995; Gullbring et al. 1998; Stempels 
\& Piskunov 2003).  The spectrum of the excess is sometimes studied by
subtracting off the spectrum of a veiled WTTS or main sequence star of
the same spectral type, while other studies use a synthetic spectrum 
computed from a model stellar atmosphere.  Here, we use a model stellar
atmosphere since we did not observe a suitable WTTS having no excess 
continuum or line emission of its own.  

The top panel of Figure 3 shows a portion of the spectrum of V2129 Oph
in the neighborhood of the \ion{He}{1} 6678 \AA\ line
with our best fit synthetic spectrum including a modest amount of veiling
($r = 0.17$, that is, a continuum excess equal to 0.17 of the local
stellar continuum) added in.  This value for the veiling is slightly higher
than previously reported values for V2129 Oph: $r = 0.0 \pm 0.1$ 
(Basri \& Batalha 1990); $r = 0.075 \pm 0.020$ (Donati et al. 2011a, though
this is only a relative veiling value and represents a lower limit 
to the true value).  For our purposes, the 
exact value of the veiling is not important.  The primary goal with such a 
fit is to predict the strength of the \ion{Fe}{1} absorption line
at 6678 \AA.  To do so, we synthesize the spectral region
between 6660 -- 6680 \AA\ as shown in Figure 4.  The two strongest features
in this region are the \ion{Fe}{1} feature at 6678 \AA\ and another
\ion{Fe}{1} feature at 6663 \AA.  Both of these features seen in Figure 4
are actualy blends of two \ion{Fe}{1} lines.  However, since all the 
components of both features are from the same element in the same ionization
stage, it should be possible to ``fit" the 6663 \AA\ feature and then
predict the strength of the 6678 \AA\ line.  We use the spectrum synthesis
code SYNTHMAG (Piskunov 1999).  The line data for this spectral region is
taken from VALD (Krupka et al. 1999, 2000).  As is often the case, the initial
predicted line strengths do not match up well with observations of the
TTSs or the Sun, so we tuned the oscillator strengths (and in a few
cases the Van der Waals broadening constants) of the strong lines until
a good match with the solar atlas (Kurucz et al. 1984) was obtained.  
In tuning the line parameters to the solar spectrum, we use the model
atmosphere and associated parameters ($v$sin$i$, macroturbulence, etc.)
for the $T_{eff} = 5731$ K scaled Kurucz model from Valenti and Piskunov 
(1996).  Once the line data is set, we then synthesize this same 
spectral region for each of our CTTSs using NextGen model atmospheres
(Allard \& Hauschildt 1995).  To do so, we must select an effective
temperature, gravity, and metallicity for each star.  We take the
gravity as log$g = 3.5$ and metallicity as [M/H] = 0.0 for all stars;
however, we pick effective temperatures from the standard NextGen grid
that are as close as possible to that for each star.  Donati et al.
(2007) adopt an effective temperature of $4500 \pm 200$ K for V2129 Oph,
and we adopt the NextGen model with $T_{eff} = 4600$ K for this star.
Yang et al. (2005) find $T_{eff} = 4126 \pm 24$ K for TW Hya, and we adopt 
the NextGen model with $T_{eff} = 4200$ K.  GQ Lup has the same spectral
type as TW Hya (K7), so we use the same NextGen model for this star as
well.  In order to do the final spectrum synthesis and fitting to the
observed profile, we must select the micro and macroturbulence as well
as the $v$sin$i$ rotational velocity.  Following Johns--Krull et al.
(1999b) we take 2.0 km s$^{-1}$ for the macroturbulent velocity and set
the microturbulence to 1.0 km s$^{-1}$.  In order to account for rotation,
we set $v$sin$i = 4.0$ km s$^{-1}$ for TW Hya (Donati et al. 2011b), 
we set $v$sin$i = 14.5$ km s$^{-1}$ for V2129 Oph (Donati et al. 2007),
and we set $v$sin$i = 6.8$ km s$^{-1}$ for GQ Lup (Guenther et al. 2005).

Using the spectrum synthesis described above, we match the strength of
the \ion{Fe}{1} feature at 6663 \AA\ and subtract the model spectrum from
the right and left circularly polarized components of the observed 
spectrum, add a pseudo-continuum of 1.0 back in,
and then follow the procedure described above for the \ion{He}{1}
5876 \AA\ line in order to measure the field in the formation region of
the 6678 \AA\ \ion{He}{1} line.  The only free parameter used to fit the 
6663 \AA\ \ion{Fe}{1} is the value of the continuum veiling.  These
veiling values and the $B_z$ values for the 6678 \AA\ emission line are
reported in Table 1.  In general our veiling values agree well with 
previously published determinations considering this quantity is often
quite variable in CTTSs.  In addition to previous veiling measurements
for V2129 Oph discussed above, veiling on TW Hya in similar spectral
regions has been shown to vary, reaching as high as 0.80 in the study
of Alencar and Batalha (2002) and as high as $\sim 0.92$ in the work
of Donati et al. (2011b).  In the case of GQ Lup, Weise et al. (2010)
found a veiling of 0.5 near the \ion{Li}{1} line at 6707 \AA, while here
we find values ranging from 0.3 to 0.6.  Again though, the exact value 
of the veiling is of secondary importance: accounting for the photospheric
absorption which is coincident with the \ion{He}{1} 6678 \AA\ emission
makes a substantial difference in the recovered field strengths for this
line.  To gauge this effect, we repeated the analysis of the 6678 \AA\ line
without making any correction for the photospheric absorption present.
Generally, the fields we measure in this case are a factor of two larger
than those reported in Table 3. 

\section{Discussion}

Examining both the null Stokes $V$ field values and the field values
obtained from the
analysis of the telluric lines shows that each of these is quite low and
generally equal to zero (as they should be) within the measured 
uncertainties.  In the case of the null spectrum, the value reaches a 
significance of $3.3\sigma$ in one case, but most values are between 
$1-2\sigma$.  For the telluric lines, all of the measured $B_z$ values
are equal to zero (as expected) to within $1\sigma$ or less.  The
uncertainties from the telluric tests are also generally lower than
the for the photospheric lines (both actual measurement and null test).
This is primarily due to the telluric lines being very sharp and strong,
allowing for very accurate measurements of any potential wavelength 
shift.  We conclude from these tests that our uncertainties in $B_z$
are generally well characterized and that our detections of polarization
in the photospheric lines and resulting measurements of $B_z$
on GQ Lup and TW Hya are real.

TW Hya has been studied with spectropolarimetry by Yang et al. (2007) and
Donati et al. (2011b).  On one of six nights, Yang et al. (2007) measured 
polarization in a dozen photospheric lines of TW Hya corresponding to a 
longitudinal magnetic field of $B_z = 149 \pm 33$ G, while finding no
significant polarization (though with larger uncertainties) on the other
five nights. Donati et al. (2011b) observed TW Hya for a total of 20 nights
and used their least squares deconvolution (LSD) analysis (Donati et al. 1997)
to measure photospheric fields ranging from $B_z = 380 - 700$ G with 
uncertainties typically of 15 G.  
Donati et al. suggest that long term temporal variability may be responsible
for the variations in $B_z$ between the measurements of Yang et al. (taken in
April 1999) and those of Donati et al. (taken in March 2008 and March 2010).  
The results presented
here, from April and May 2010, agree well with those of Yang et al. and
less so with those of Donati et al. (2011b).

Yang et al. (2007) measured polarization in the \ion{He}{1} 5876 \AA\ and
6678 \AA\ emission lines of TW Hya, as well as in the \ion{Ca}{2}
8498 \AA\ emission line.  Yang et al. report possible rotational modulation
of the polarization in the \ion{He}{1} 5876 \AA\ line, with implied fields
varying from $\sim -1450$ to $\sim -1800$ G.  
Donati et al. (2011b) also measure strong 
polarization in the \ion{He}{1} 5876 \AA\ line with variations again 
suggestive of rotational modulation.  The field strength implied in this case
ranges from $\sim -2000$ to $\sim -3500$ G.  Donati et al. attribute the
significant difference between their \ion{He}{1} measurements and those of 
Yang et al.  to perceived differences
in the way the fields were measured in the two studies.  Donati et al.
measured the field only in the narrow component of the line for similar
reasons to those cited above in \S 3.3, and they state that Yang et al.
(2007) used the entire \ion{He}{1} line in their field determination;
however, this is not correct.  Yang et al. (2007) also focussed only on
the narrow component of the \ion{He}{1} emission line in their field
determinations (Yang, private communication).  The field measurements 
reported here in Table 3 for $B_z$ in the \ion{He}{1} 5876 \AA\ emission
line generally agree well with those of Yang et al. and are significantly
less than most of the fields reported by Donati et al.  However, Yang et al. 
used a cross correlation technique to measure line shifts and resulting field
strengths, while here we use the center of gravity technique to measure 
line shifts.  

In order to verify that the difference in measurement technique does not 
introduce a spurious difference in the recovered field, we reanalyzed the
Yang et al. (2007) spectrum of TW Hya obtained on 21 April 1999 (the 
strongest \ion{He}{1} field they found) using the same center of gravity
technique employed above.  Briefly, this spectrum was obtained with the
2.7 m Harlan J. Smith telescope of McDonald Observatory used to feed the
Robert G. Tull coude echelle spectrometer (Tull et al. 1995).  The starlight
was split into its circularly polarized components using a Zeeman 
Analyzer (ZA, Vogt et al. 1980) placed in front of the spectrometer
slit.  The ZA contains a Babinet-Soleil phase compensator used to correct for
a potential reduction in sensitivity to circular polarization which can be 
introduced by the non-azimuthally symmetric reflections in the coude mirror 
train.  Such bounces can introduce some linear polarization into an originally
circularly polarized beam and the phase compensator is used to convert this
linear polarization back into the original circular polarized signal (see
Vogt et al. 1980 and Vogt 1978).  The exposure of
TW Hya for that night totalled 4300 s.  Yang et al. (2007) measured a
field of $-1806 \pm 114$ G for the narrow component of the \ion{He}{1} 5876
\AA\ line using a cross correlation analysis, while we find a field of 
$-2326 \pm 118$ G using the center of gravity technique, fully consistent
with our current measurements.  This measurement of the field is stronger by 
$520 \pm 164$ G than that of Yang et al. (2007), representing a $3.2\sigma$
difference.  The difference is driven both by the method of estimating
the line shift (center of gravity versus cross correlation) as well as in
the treatment of removing the broad component.  
As a result, it does not appear that
differences in the measurement technique can account for the variations in
the field strength in the \ion{He}{1} emission line formation region found
between Donati et al. (2011b) and this study plus that of Yang et al. (2007).
It is more likely that intrinsic variability is at work in this
accretion diagnostic, and indeed,
Donati et al. (2011b) find significant differences in the \ion{He}{1} field
strength from one rotation phase to the next in TW Hya, while the bulk of
their field measurements are larger in magnitude than either Yang et al.
(2007) or those here.

One of the motivations for this study was to verify and expand on the 
intriguing result that the field measured from the \ion{He}{1} 6678 \AA\ line
is significantly different (stronger) than that measured in the \ion{He}{1}
5876 \AA\ line (e.g. Donati et al. 2008).  In principle, measuring the field
in different emission line diagnostics could offer a means for probing the
magnetic field structure through the accretion shock on CTTSs.  As described
in \S 3.3, we discovered that there is a strong photospheric \ion{Fe}{1}
absorption line coincident in wavelength with the \ion{He}{1} 6678 \AA\ line
that can severely affect the field strength measured in this emission line
if the photospheric line is not properly accounted for.  As described above,
we attempted to correct
for the \ion{Fe}{1} photospheric line by computing veiled 
model spectra for each of our observations and subtracted the resulting
model from the observations before measuring the field.  Doing so produced
field measurements in the \ion{He}{1} 6678 \AA\ line that are very 
consistent with those measured in the 5876 \AA\ line of the same element.
The field strengths recovered from the two lines are the same to within
3$\sigma$ for all observations.
If we do not account the \ion{Fe}{1} photospheric
absorption line, we generally recover a field in the 6678 \AA\ line that
is twice as strong as reported in Table 3.  In a separate test, we corrected
for the photospheric absorption line in the observations of TW Hya and GQ Lup
using a veiled version of the observed V2129 Oph spectrum.  While this is
not ideal due to the weak \ion{He}{1} 6678 \AA\ emission from V2129 Oph,
the results were the same as those in Table 3 within the uncertainties.  As
a result, it appears that the field in the two \ion{He}{1} lines is 
essentially identical, at least in TW Hya and GQ Lup, but that special
care must be taken when analyzing the 6678 \AA\ line.  
We also note from earlier that our analysis of the
5876 \AA\ line assumed it formed in the complete Paschen-Back regime (i.e.
$g_{eff} = 1.0$).  Since shock models suggest the field strength in these
two lines should essentially be the same, and that is indeed what we find,
these results also suggest that our treatment of the 5876 \AA\ line in
the complete Paschen-Back effect is appropriate, at least for the strong
magnetic fields recovered here. 

The magnetic field on GQ Lup appears to be quite remarkable.  The field
detected in the photospheric absorption lines is fairly typical in magnitude
($|B_z| \sim 200$ G) relative to the strength detected on many CTTSs.  
However, the field ($\sim 6$ kG) detected in the \ion{He}{1} line formation
region, the accretion shock, is substantially stronger in magnitude than that
observed in most CTTSs.  Symington et al. (2005) reported some quite strong
fields in this line from a few CTTSs; however, also with substantial error
bars.  Higher precision measurements (with uncertainties from $\sim 0.1 - 0.4$
kG) have peak fields measured in the \ion{He}{1} 5876 \AA\ line of 
$|B_z| < 3.4$ kG (Donati et al. 2011) with peak values typically 
$|B_z| \sim 2$ kG depending on the specific star in question (Johns--Krull
et al. 1999a; Valenti \& Johns--Krull 2004; Yang et al. 2005; Donati et 
al. 2007, 2008, 2010b).  Very recently, Donati et al. (2012) present 
polarimetric measurements of GQ Lup at two different epochs (July 2009,
June 2011).  While the photospheric fields they detect do show some variation
from one epoch to the other, our measurements over about half the rotation
period found by Donati et al. are consistent with the fields found at
both epochs.  On the ther hand, Donati et al. find a substantial decrease in
the large scale field controlling the accretion over the two years between
their epochs.  Our data (April 2010) is fully consistent with their earlier
epoch, with our \ion{He}{1} $B_z$ determinations matching the strongest
values they observe in July 2009, providing additional constraints on
the timescale involved in the magnetic field change observed by Donati
et al.

\acknowledgements

We wish to thank ESO staff in Santiago and on La Silla for their hospitality
and help during the observing run there.  This research has made use of the
Simbad Astronomical database, the VALD line database, and the NASA 
Astrophysics Data System.  CMJ-K wishes to acknowledge partial support
for this research from the NASA Origins of Solar Systems program through 
grant numbers NNX10AI53G made to Rice University.  OK is a Royal Swedish 
Academy of Sciences Research Fellow, supported by grants from Knut and Alice 
Wallenberg Foundation and Swedish Research Council.  Finally, we thank an
anonymous referee for helpful suggestions to improve the presentation in
the manuscript.

\clearpage
 
\begin{deluxetable}{lccccccccc}
\tablewidth{12.5truecm}   
\tablecaption{Observing Log\label{obslog}}
\tablehead{
   \colhead{}&
   \colhead{UT}&
   \colhead{}&
   \colhead{$S/N$\tablenotemark{b}}&
   \colhead{$S/N$\tablenotemark{b}}&
   \colhead{\ion{He}{1} 5876\AA}&
   \colhead{\ion{He}{1} 5876\AA}&
   \colhead{\ion{He}{1} 6678\AA}&
   \colhead{\ion{He}{1} 6678\AA}&
   \colhead{}\\[0.2ex]
   \colhead{UT Date}&
   \colhead{Time\tablenotemark{a}}&
   \colhead{Star}&
   \colhead{5876\AA }&
   \colhead{6678\AA }&
   \colhead{NC $W_{eq}$ (\AA )}&
   \colhead{BC $W_{eq}$ (\AA )}&
   \colhead{NC $W_{eq}$ (\AA )}&
   \colhead{BC $W_{eq}$ (\AA )}&
   \colhead{$r$\tablenotemark{c}}
}
\startdata
29 April 2010 & 1:00 & TW Hya & 59 & 56 & $1.110 \pm 0.004$ & $1.179 \pm 0.007$& $0.387 \pm 0.003$ & $0.269 \pm 0.005$ & 1.00 \\
 &  4:16 & GQ Lup & 42 & 42 & $0.604 \pm 0.005$ & $0.137 \pm 0.008$ & $0.129 \pm 0.003$ & $0.030 \pm 0.005$ & 0.40 \\
30 April 2010  & 0:56 & TW Hya & 66 & 78 & $1.282 \pm 0.004$ & $2.037 \pm 0.008$ & $0.363 \pm 0.002$ & $0.479 \pm 0.004$ & 1.40 \\
  & 3:02 & GQ Lup & 31 & 29 & $0.530 \pm 0.005$ & $0.235 \pm 0.009$ & $0.175 \pm 0.005$ & \nodata & 0.65 \\
  & 7:28 & V2129 Oph\tablenotemark{d} & 65 & 68 & $0.134 \pm 0.002$ & \nodata & \nodata & \nodata & 0.175 \\
02 May 2010  & 3:09 & TW Hya & 90 & 82 & $1.312 \pm 0.003$ & $1.307 \pm 0.005$ & $0.393 \pm 0.011$\tablenotemark{e} & $0.281 \pm 0.011$ & 1.00 \\
  & 6:35 & GQ Lup & 83 & 72 & $0.365 \pm 0.002$ & $0.620 \pm 0.020$\tablenotemark{f} & $0.037 \pm 0.006$ & $0.110 \pm 0.007$ & 0.30 \\
\enddata
\tablenotetext{a}{This is the midpoint of the 4 $\times$ 1800 s exposures that
make up each total observation.}
\tablenotetext{b}{This is the $S/N$ in the continuum near the respective 
\ion{He}{1} emission lines, calculated from the final Stokes $I$ spectrum.}
\tablenotetext{c}{This is the veiling in the vicinity of the 6678 \AA\ \ion{He}{1} line.}
\tablenotetext{d}{The entries for V2129 Oph which show no data are due to either
there being no clear broad component in the case of the 5876 \AA\ line, or 
there being no emission above the continuum in the case of the 6678 \AA\ line.
As discussed later in the text, there is some filling in of a nearby 
photospheric absorption line by \ion{He}{1} emission at 6678 \AA ; however,
the overall line remains below the continuum level and so we do not record
and emission equivalent width here.}
\tablenotetext{e}{Due to apparent photospheric absorption on the blue side of
the line, there is some ambiguity in how to separate the NC and the BC on
this side of the line profile.  The reported value and larger uncertainty here
takes into account repeated measurements (averaging to get the value) where 
more or less of the emission is attributed to the BC or the NC.}
\tablenotetext{f}{The line on this night appeared to have quite extended BC
wings, making it difficult to establish exactly where the line rejoined the
continuum.  The measurement and uncertainty are formed by averaging 
conservative and more broadly inclusive measurements of the 5876 \AA\ line
for this night.}
\end{deluxetable}

\clearpage

\begin{deluxetable}{lcc}
\tablewidth{7.5truecm}   
\tablecaption{Lines Used for Photospheric Field Analysis\label{photlines}}
\tablehead{
   \colhead{Element}&
   \colhead{Wavelength (\AA )}&
   \colhead{Land\'e-$g_{eff}$}
}
\startdata
\ion{V}{1} & 6058.139 & 2.14 \\
\ion{Ti}{1} & 6064.626 & 1.99 \\
\ion{Fe}{1} & 6173.334 & 2.50 \\
Blend & 6216.355   &    1.59 \\
\ion{Fe}{1} &   6219.278  &     1.66 \\
\ion{Fe}{1}\tablenotemark{d} & 6232.640 & 1.99 \\
\ion{Fe}{1} &   6246.316  &     1.58 \\
\ion{V}{1} &   6251.827   &    1.57 \\
\ion{Fe}{1} &   6252.554  &    0.95 \\
\ion{V}{1} &   6274.648   &    1.53 \\
\ion{V}{1} &   6285.149   &    1.58 \\
\ion{Cr}{1} & 6330.091 & 1.83 \\
\ion{Fe}{1} & 6336.823 & 2.00 \\
\ion{Fe}{1} &   6393.600  &    0.91 \\
\ion{Fe}{1} &   6408.018  &     1.01 \\
\ion{Fe}{1} &   6411.647  &     1.18 \\
\ion{Fe}{1} &   6421.349  &     1.50 \\
\ion{Ca}{1}\tablenotemark{a} &   6439.075  &     1.12 \\
Blend & 6462.629   &   0.98 \\
\ion{Ca}{1} &   6471.662  &     1.20 \\
\ion{Fe}{1}\tablenotemark{b} & 6475.624 & 1.90 \\
\ion{Fe}{1} &   6481.869  &     1.50 \\
\ion{Ca}{1} &   6493.780  &    0.88 \\
\ion{Fe}{1} &   6498.937  &     1.38 \\
\ion{Ca}{1}\tablenotemark{c} &   6499.649  &    0.96 \\
\ion{Fe}{1} &   6518.365  &     1.15 \\
\ion{V}{1} &   6531.415   &    1.57 \\
\ion{Cr}{1}\tablenotemark{a} & 6537.921 & 1.71 \\
\ion{Fe}{1} &   6574.227  &     1.25 \\
\ion{Ni}{1} &   6586.308  &     1.02 \\
\ion{Fe}{1} &   6593.870  &     1.13 \\
\ion{Ti}{1} &   6599.105  &    0.98 \\
\ion{V}{1} &   6624.838   &    1.43 \\
\ion{Ni}{1} &   6643.628  &     1.31 \\
\ion{Fe}{1} &   6663.334  &     1.53 \\
\ion{Li}{1} &   6707.799  &     1.25 \\
\ion{Fe}{1} & 6710.316 & 1.69 \\
\ion{Ca}{1} &   6717.681  &     1.01 \\
\ion{Ti}{1} &   6743.122  &     1.01 \\
\ion{Fe}{1} &   6750.149  &     1.50 \\
\enddata
\tablenotetext{a}{This line excluded from all TW Hya analysis due to apparent
additional blending.}
\tablenotetext{b}{This line excluded from analysis of GQ Lup on 30 April 2010
and from analysis of V2129 Oph due to significant cosmic ray hit.}
\tablenotetext{c}{This line excluded from analysis of GQ Lup on 02 May 2010
due to significant cosmic ray hit.}
\tablenotetext{d}{This line excluded from analysis of V2129 Oph
due to significant cosmic ray hit.}
\end{deluxetable}

\clearpage

\begin{deluxetable}{lcccccc}
\tablecaption{Magnetic Field Measurements\label{fieldmeasure}}
\tablehead{
   \colhead{}&
   \colhead{}&
   \colhead{$B_Z$ (Phot)}&
   \colhead{$B_Z$ (Null)}&
   \colhead{$B_Z$ (Tel.)}&
   \colhead{$B_Z$ (\ion{He}{1} 5876\AA )}&
   \colhead{$B_Z$ (\ion{He}{1} 6678\AA )}\\[0.2ex]
   \colhead{UT Date}&
   \colhead{Star}&
   \colhead{(G)}&
   \colhead{(G)}&
   \colhead{(G)}&
   \colhead{(kG)}&
   \colhead{(kG)}
}
\startdata
29 April 2010  & TW Hya & $181 \pm 27$ & $8 \pm 27$ & $-3 \pm 7$ & $-2.38 \pm 0.10$ & $-2.46 \pm 0.13$ \\
  & GQ Lup & $-195 \pm 35$ & $0 \pm 35$ & $-1 \pm 11$ & $6.05 \pm 0.24$ & $6.15 \pm 0.39$ \\
30 April 2010  & TW Hya & $176 \pm 20$ & $-15 \pm 20$ & $2 \pm 5$ & $-1.87 \pm 0.09$ & $-2.14 \pm 0.11$ \\
  & GQ Lup & $-236 \pm 49$ & $-52 \pm 49$ & $-4 \pm 14$ & $5.63 \pm 0.39$ & $6.39 \pm 0.50$ \\
  & V2129 Oph & $-19 \pm 32$ & $-54 \pm 31$ & $-8 \pm 8$ & $2.87 \pm 0.39$ & $3.10 \pm 0.56$ \\
02 May 2010  & TW Hya & $180 \pm 16$ & $-13 \pm 16$ & $0 \pm 6$ & $-2.25 \pm 0.09$ & $-2.53 \pm 0.08$ \\
  & GQ Lup & $-81 \pm 18$ & $6 \pm 18$ & $-4 \pm 6$ & $5.23 \pm 0.18$ & $4.87 \pm 0.31$ \\
\enddata
\end{deluxetable}

\clearpage

\begin{figure} 
\epsscale{.70}
\plotone{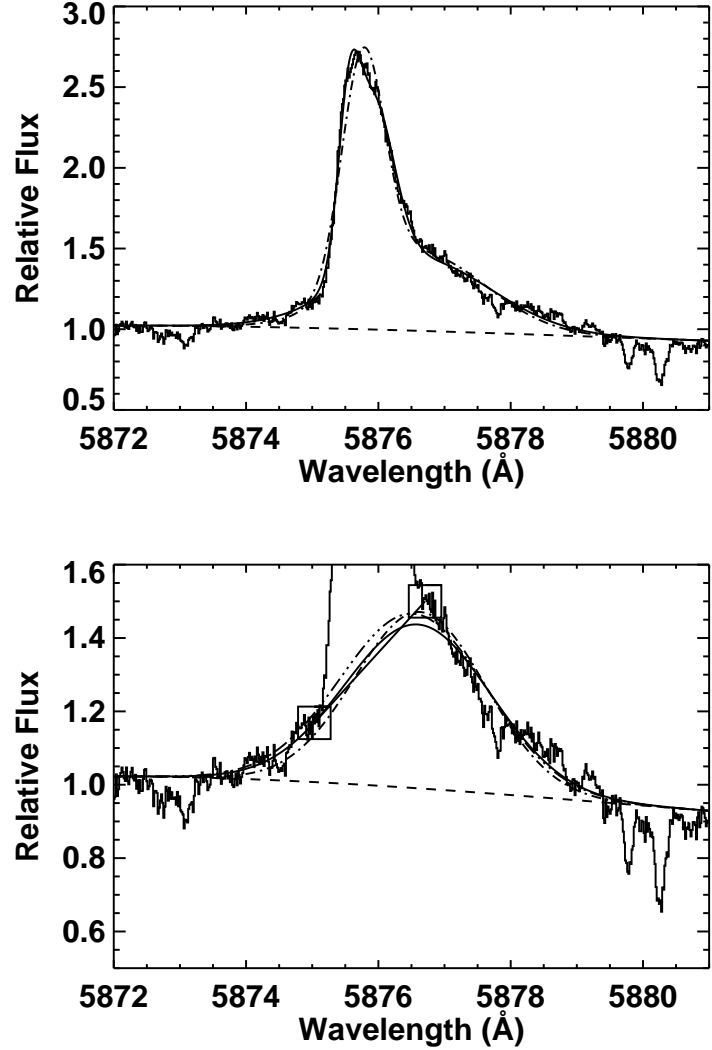}
\caption{In each panel, the continuum normalized \ion{He}{1} 5876 \AA\ line
profile of TW Hya from 29 April 2010 is shown in the solid histogram.  The top
panel shows two multi Gaussian fits to the profile, with a two Gaussian
fit shown in the dash-dot line and a 3 Gaussian fit shown in the smooth
solid line.  The bottom panel zooms in on the line to show the recovered BC
profiles.  The BC from the 3 Gaussian fit is shown in the smooth solid line
and that from the two Gaussian fit is shown in the dash-dot line.  The
dash-triple dot line shows a single Gaussian fit following Batalha et al. 
(1996).  The solid straight line connecting the two large squares shows a by
eye estimate of the point on both the blue and red side of the line (as seen
in Stokes $I$) where the NC and BC join with a linear interpolation between
these points to define the separation of the NC and the BC.
}
\end{figure}

\clearpage
\begin{figure} 
\epsscale{.90}
\plotone{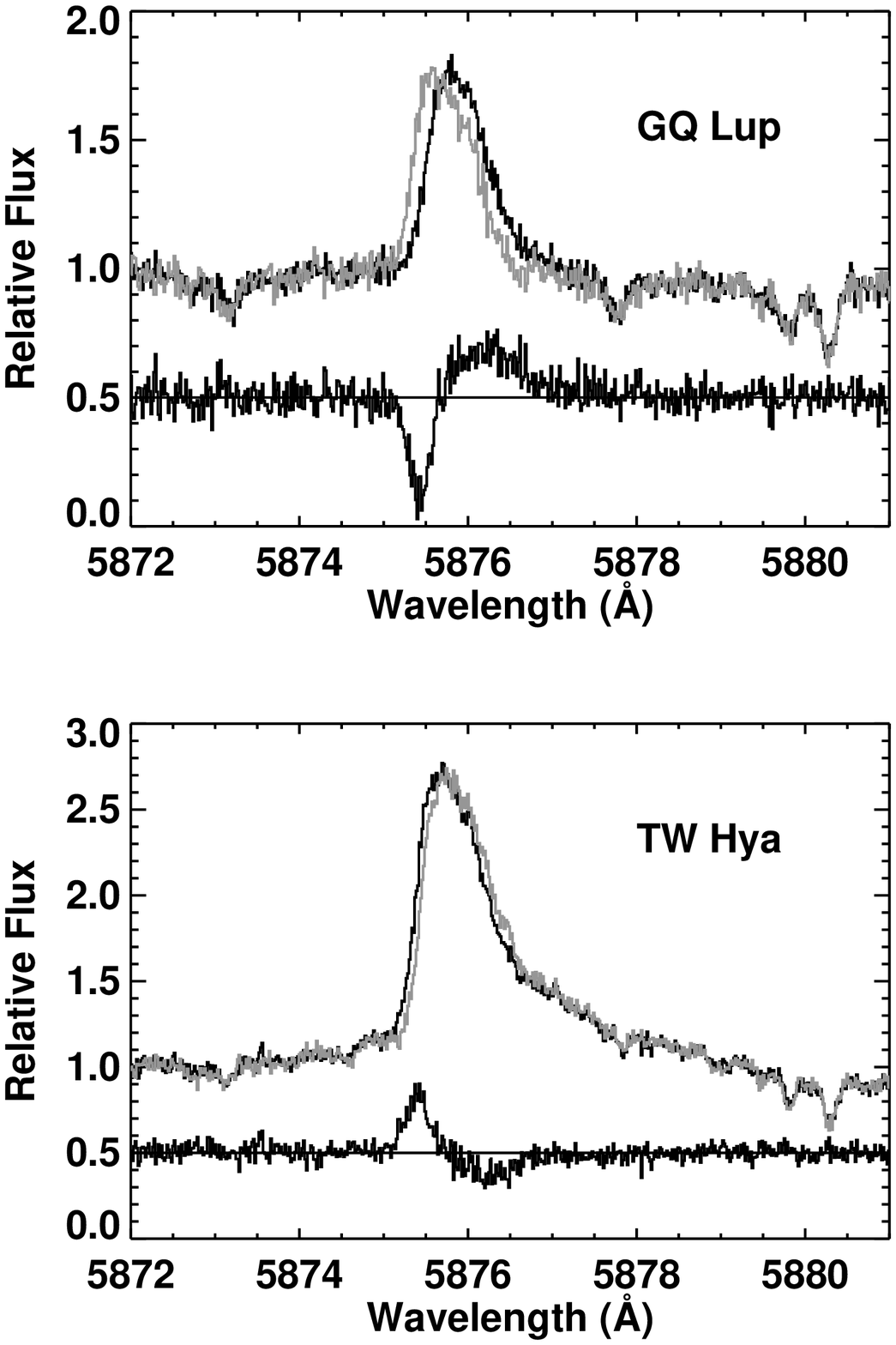}
\caption{In each panel, the continuum normalized right circularly 
polarized component of the \ion{He}{1} 5876 \AA\ line profile is shown
in the upper black curve and the left circularly polarized component is 
shown in the gray curve.  The difference (right -- left) of the two,
offset by 0.5, is shown in the lower black curve.  This difference is
$2 \times $ Stokes $V$ with the profiles normalized in this way.
}
\end{figure}

\clearpage

\begin{figure} 
\plotone{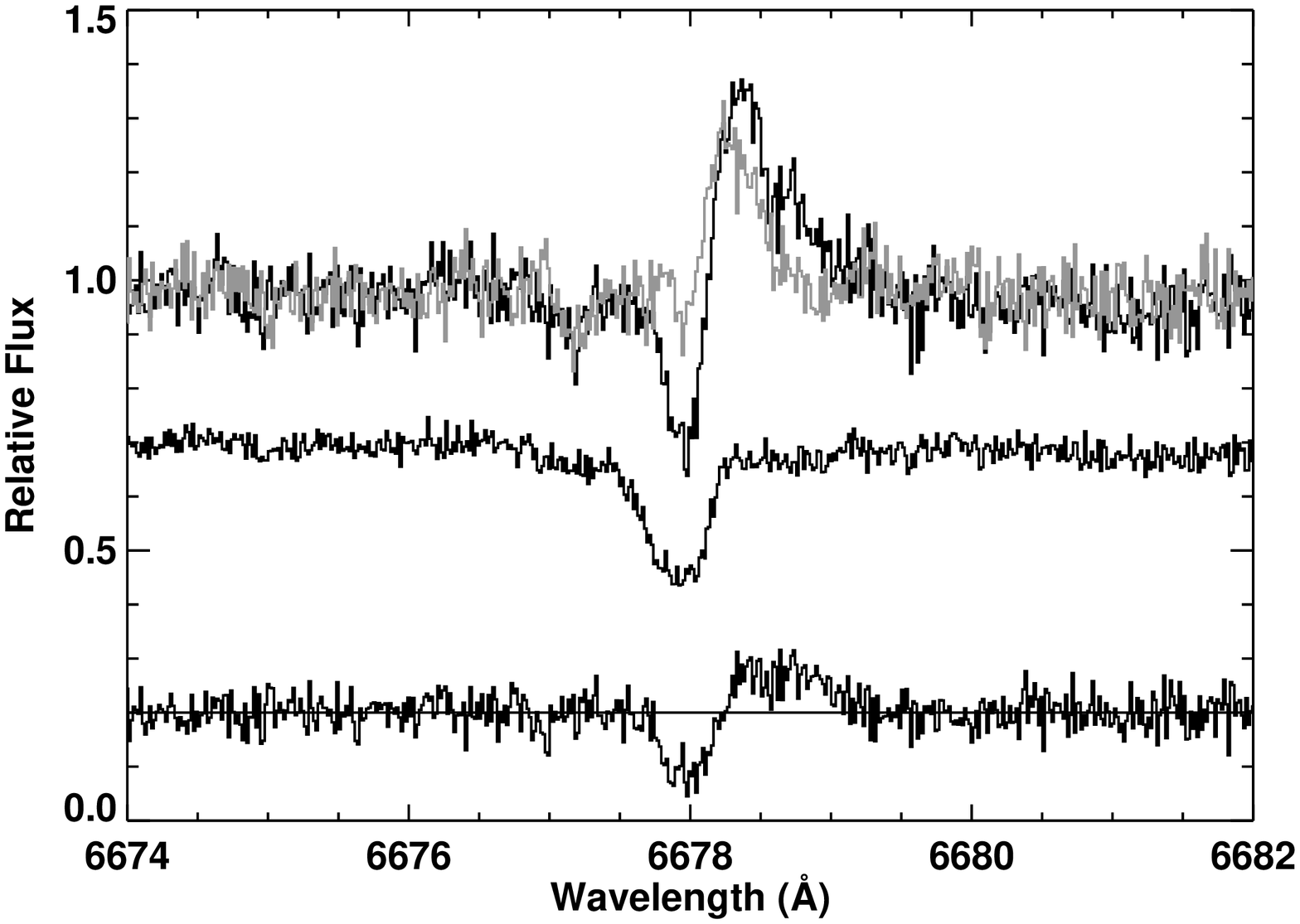}
\caption{The continuum normalized right circularly 
polarized component of the \ion{He}{1} 6678 \AA\ line profile of GQ Lup is
shown in the upper black curve and the left circularly polarized component is 
shown in the gray curve.  The difference (right -- left) of the two,
offset by 0.2, is shown in the bottom black curve.  This difference is
$2 \times $ Stokes $V$ with the profiles normalized in this way.  The middle
black curve shows the same region in the Stokes $I$ spectrum of V2129 Oph,
revealing the presence of a strong \ion{Fe}{1} photospheric absorption line
which can alter the inferred field in the He formation region if the Fe
line is not properly accounted for.
}
\end{figure}

\clearpage

\begin{figure} 
\epsscale{.70}
\plotone{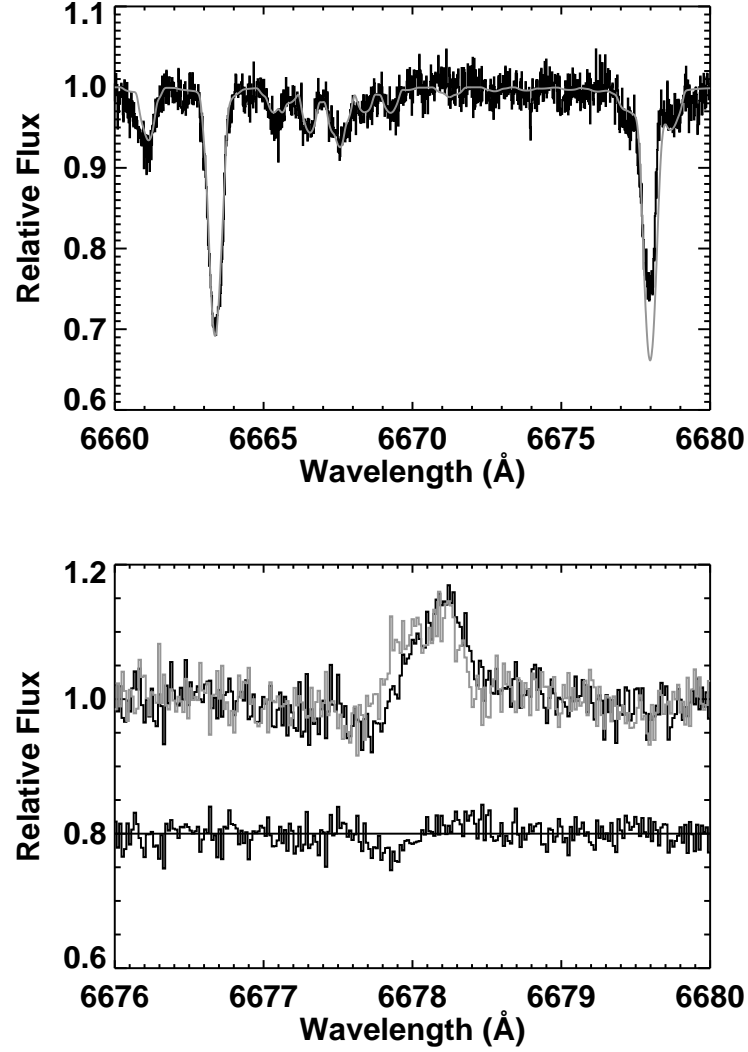}
\caption{The upper panel shows the observed Stokes $I$ spectrum of V2129
Oph in an expanded region near the \ion{He}{1} 6678 \AA\ line.  The two
strongest lines are \ion{Fe}{1} lines at 6678 \AA\ and 6663 \AA.  The
smooth gray curve in the upper panel shows a synthetic fit to the spectrum,
showing that there is excess emission the spectrum of V2129 Oph at 6678 \AA\
due to a weak \ion{He}{1} emission line superimposed on the photospheric
absorption spectrum.  In the bottom panel, the upper black
curve shows the continuum normalized right circularly polarized component of 
the \ion{He}{1} 6678 \AA\ line profile of V2129 Oph after subtracting off the
synthetic fit and adding 1.0 back to the result.  The gray curve shows the
same for the left circularly polarized component of the spectrum.
The difference (right -- left) of the two, offset by 0.8, is shown in the 
lower black curve.  This difference is $2 \times $ Stokes $V$ with the 
profiles normalized in this way.
}
\end{figure}

\end{document}